\documentclass[aps,amsfonts,amsmath,prd,preprint,nofootinbib]{revtex4}
\pdfoutput=1
\usepackage{graphicx}

\newcommand{\beq}{\begin{equation}}
\newcommand{\eeq}{\end{equation}}

\def\lap{\lower.5ex\hbox{$\; \buildrel < \over \sim \;$}}
\def\gap{\lower.5ex\hbox{$\; \buildrel > \over \sim \;$}}

\def\be{\begin{equation}}
\def\ee{\end{equation}}
\def\ba{\begin{eqnarray}}
\def\ea{\end{eqnarray}}

\begin{document}

\title{Fireballs from Superconducting Cosmic Strings}

\author{Andrei Gruzinov$^{a}$ and  Alexander Vilenkin$^b$}

\address{
$^a$ CCPP, Department of Physics, New York University,
New York, NY 10001\\
$^b$Institute of Cosmology, Department of Physics and Astronomy, Tufts University,
 Medford, MA 02155}

\begin{abstract}

Thermalized fireballs should be created by cusp events on superconducting cosmic strings. This simple notion allows to reliably estimate particle emission from the cusps in a given background magnetic field.  With plausible assumptions about intergalactic magnetic fields, the cusp events can produce observable fluxes of high-energy photons and neutrinos with unique signatures.

\end{abstract}

\maketitle

\section{Introduction}

Cosmic strings are linear defects that could be formed at a symmetry breaking phase transition in the early universe (for a review and references see \cite{Book}).  Strings predicted in most grand unified models respond to external electromagnetic fields as thin superconducting wires \cite{Witten}.  As they move through cosmic magnetic fields, such strings develop electric currents.  Oscillating loops of superconducting string emit short bursts of highly beamed electromagnetic radiation \cite{VV87,Spergel87,Babul87} and high-energy particles \cite{Barr87,BOSV,Tanmay}.

The bursts originate at short string segments, called cusps, where the string briefly develops a large Lorentz factor.  Even though particle emission from cusps has been extensively studied in the literature, it appears that some important aspects of this process have not been adequately described.  The ejected particles and their decay products were treated individually, disregarding their interaction with one another.  And the electromagnetic fields produced by the string were treated classically, disregarding the possibility of pair production.  Here, we point out that in a cusp event a large energy is injected in a very small volume.  This energy thermalizes, resulting in a relativistic fireball.  The fireball then expands and cools, until it becomes optically thin, at which point the particles freely disperse.  The fireball is boosted by the large Lorentz factor $\gamma$ of the cusp, so it appears to an observer as a narrow jet with an opening angle $\sim \gamma^{-1}$.  This is similar to earlier treatments, but the particle composition of the jet is now very different.  The predicted observational effects of superconducting strings should therefore be reconsidered.

We begin in the next section with a brief review of cosmic string properties and evolution. Then we describe the cusp formation and emission from cusps (\S\ref{CoL}), thermalization of the cusp emission (\S\ref{FfC}), the fireball expansion (\S\ref{FE}), and finally the resulting observables (\S\ref{Obs}).

\section{String properties and evolution}

Numerical simulations of string evolution indicate that strings evolve in a self-similar manner.  A Hubble-size volume at any time $t$ contains a few long strings stretching across the volume and a large number of closed loops of length $l\ll t$ (for an up to date review and references, see \cite{BPOS}).  The loops oscillate periodically and lose their energy, mostly by gravitational radiation.  For a loop of length $l$, the oscillation period is $l/2$ and the lifetime is
\beq
\tau_l \sim l/k_g G\mu.  
\eeq
Here, $k_g\sim 50$ is a numerical coefficient, $G$ is Newton's constant, and $\mu$ is the string mass per unit length, which is related to the symmetry breaking energy scale $\eta$ as
\beq
\mu \sim \eta^2.
\eeq

Gravitational waves emitted by loops over the cosmic history add up to a stochastic gravitational wave background with a wide spectrum of frequencies.  Requiring that the predicted amplitude of this background is not in conflict with the millisecond pulsar observations, one can impose an upper bound on the string mass parameter $\mu$: $G\mu \lesssim 3\times 10^{-9}$ \cite{BPOS}.  The corresponding bound on $\eta$ is 
\beq
\eta_{14}\lesssim 5,
\eeq
where $\eta_{14} = \eta/10^{14} GeV$.

The most interesting for our discussion here are the most numerous loops surviving at the preset time $t_0$.  These loops were formed in the radiation era and have lifetimes $\tau_l \sim t_0$.  Their length is 
\beq
l\sim 50 G\mu t_0 \sim 10^{20}{\rm cm}~ \eta_{14}^2
\eeq
and their number density is
\beq
n\sim 10^{-5} (G\mu)^{-3/2} t_0^{-3}.
\eeq
There are 
\beq
N \sim 10^{10}\eta_{14}^{-3}
\eeq
such loops within the present horizon.

\section{Cusps on Loops}\label{CoL}

At the time of their formation, cosmic string loops consist of nearly straight segments separated by sharp kinks \cite{BPO16}.  Such loops do not develop cusps.  However, the kinks are gradually smoothed by the gravitational back-reaction and all but disappear towards the end of the loop's life.  The most numerous loops, decaying at the present epoch, are expected to be very smooth and to
develop cusps during each oscillation period \cite{BPO16}.  Cusp events occur on  such loops every 
\beq
t_c \sim 3\times 10^{9}{\rm s}~ \eta_{14}^2.
\eeq

A string loop of length $l$ oscillating in a magnetic field $B$ acquires an electric current
\beq
J_0 \sim 0.1 e^2 Bl.
\eeq
The current is strongly enhanced at cusp events, resulting in powerful bursts of EM radiation.  In the vicinity of a cusp, a string segment of initial length $l/\gamma$ shrinks by a factor of $\gamma$, so its length becomes 
\beq
l_\gamma \sim \gamma ^{-2}l,
\eeq
and develops a Lorentz factor $\gamma$.  In the local rest frame of the string, the current is enhanced by the same factor, and an EM pulse of energy 
\beq \label{Ec}
E_c \sim 2\times 10^{37}{\rm erg}~ B_{-9}^2\eta_{14}^6
\eeq
is released into a region of characteristic size $l_\gamma$ during the time $\sim l_\gamma$.\footnote{Eq.~(\ref{Ec}) follows from the angular distribution of the EM energy in the lab system derived in \cite{VV87,Spergel87}.  The energy emitted within angle $\theta$ of the string velocity at the cusp is ${E}_\theta \sim 10 J_0^2 l /\theta$.  It comes from a segment with $\gamma\sim 1/\theta$.  Transforming to the frame of the segment, we have $E_c\sim 10 J_0^2 l$, independent of $\gamma$, which gives Eq.~(\ref{Ec}).}

The Lorentz factor $\gamma$ increases towards the central point of the cusp, terminating at   
\beq\label{Gac}
\gamma _c \sim 2\times 10^{10}~ B_{-9}^{-1}\eta_{14}^{-1}.
\eeq
This limiting value of $\gamma$ is due to the inertia of the charge carriers on the string \cite{BPOV,BPO}.

Charged particles moving in opposite directions along the string can scatter off one another and escape from the string \cite{Barr87}.  This process can be very efficient in the cusp region, where the particle density is very high.  However, currents flowing in opposite directions are not expected to be balanced, so ejection of particles does not change the order of magnitude of the current, and the electromagnetic burst continues to develop with the Lorentz factor growing up to $\gamma_c$.  
The ejected particles scatter off one another and pair-create in the
strong electromagnetic field near the cusp.\footnote{Interactions of
  the ejected particles with the EM field of the string have been
  discussed in Ref.~\cite{Rubinstein}.}  They quickly thermalize and form a relativistic fireball.  This alternative mechanism of fireball formation is sensitive to the string microphysics. For some parameter values the energy carried away by the ejected particles can be comparable to the electromagnetic energy \cite{BOSV}, but it can also be much smaller.  Here we shall focus on fireballs of EM origin.

The proper energy Eq.(\ref{Ec}) is negligibly small for an object at a  cosmological distance, but the Lorentz boost Eq.(\ref{Gac}) is very large. In the laboratory frame (the CMB frame): (i) the energy of the EM pulse is Lorentz boosted, (ii) its length scale is Lorentz contracted, (iii) the pulse is beamed into an opening angle $\sim \gamma ^{-1}$. Altogether, this gives an equivalent isotropic luminosity
\beq \label{iso}
L_{\rm iso} \sim 4\times 10^{89}\frac{\rm erg}{\rm s}~ B_{-9}^{-4}\eta_{14}^{-2}\left( \frac{\gamma}{\gamma _c} \right)^6
\eeq
-- unheard of in astrophysics. However, as the beaming angle is small, the probability to get illuminated by the most energetic part of cusp emission, with $\gamma \sim \gamma _c$, is negligibly small.  

Cusp events might become observable in two ways. Emission into larger beaming angles does occur at smaller $\gamma$.  Also, emission initially beamed into a small solid angle can spread out by propagation through the intergalactic medium (IGM). Both of these possibilities are considered in the following sections. Here we will show that astrophysically  interesting fluences and fluxes can arise in both scenarios, so that further investigation might be worthwhile. 

First consider cusp events that illuminate a given observer once per time $t=t_y$ years. This requires beaming angles $\theta \sim \gamma ^{-1}$ given by
\beq
t \sim \frac {t_c}{N}\theta ^{-2}.
\eeq
Knowing $\gamma$, we estimate the  fluence (for an event at a cosmological distance $r\sim 3$ Gpc),  $f\sim \gamma ^3 E_c/r^2$, or
\beq \label{flue}
f \sim 2\times 10^{-7}\frac{\rm erg}{\rm cm ^2}~ B_{-9}^{2}\eta_{14}^{-3/2}t_y^{3/2}.
\eeq

 Another possibility is that a fair fraction of the cusp energy, initially narrowly beamed, spreads out without great losses, where "losses" mean photons of energy $\lesssim 1$GeV which would be unobservable on the diffuse gamma-ray background. Then the figure of merit is the time averaged flux, $F\sim \frac{N\gamma _cE_c}{t_cr^2}$, giving
 \beq \label{flux}
F \sim  10^{-8}\frac{\rm erg}{\rm cm ^2\cdot s}~ B_{-9}.
\eeq

Intergalactic magnetic field is likely inhomogeneous, with different magnetic field values $\sim B$ occupying different volumetric fractions of the universe $f_B$. Then the fluence Eq.(\ref{flue}) is multiplied by $f_B^{3/2}$, the flux Eq.(\ref{flux}) is multiplied by $f_B$, and the results are summed over all $B$ values.

Throughout this paper we assume that cosmic string loops lose their energy mostly by gravitational radiation.  This holds for $\eta_{14}\gtrsim 10^{-4}f_B B_{-9}$, while for smaller values of $\eta$ the EM radiation is the dominant energy loss mechanism.  In the latter case, the lifetime of the loops is independent of their size and all loops formed in the radiation era decay before the present epoch \cite{BOSV}.

We have only considered the cusp events at redshifts $z\lesssim 1$.  The analysis can be extended to include higher redshifts \cite{BHV,BOSV}.  One finds that the wait time for cusp events at $z\gtrsim 1$ with a given energy fluence $f = 10^{-8} f_{-8} ~erg/cm^2$ is
\beq
t_y \sim 10^{-2} \eta_{14} f_{-8}^{2/3}f_B^{-1}(z)B_{-9}^{-4/3}(z)(1+z)^{7/4}(\sqrt{1+z}-1)^{-2/3}.
\eeq
This shows that for a uniform primordial magnetic field with $f_B(z) = 1$ and $B(z)\propto (1+z)^2$, the contribution of high redshift events could be significant.  On the other hand, if the intergalactic fields originated during and after the galaxy formation epoch, most of the detectable cusp events are likely to occur at $z\sim 1$. 

We see that, unlike $L_{\rm iso}$, both the fluence $f$ and the flux $F$ have reasonable values. They are large enough to be detectable, but not too large to be immediately ruled out. 

\section{Fireballs from Cusps}\label{FfC}
 
 Observability of fluences and fluxes Eqs.(\ref{flue},\ref{flux}) depends on which particle (photon or neutrino) and at what energy reaches Earth. We must then describe the evolution of the emitted EM pulse. Here we show that for large enough $\gamma$, the EM pulse thermalizes, producing a fireball of initial temperature $T_f\gtrsim 0.1$ MeV. This allows to predict what gets injected into the ISM and then  also what reaches Earth. 
   
 The idea that large energy, quickly released into a small volume, thermalizes is nothing new in astrophysics.  In the context of GRBs (gamma ray bursts) it has been discussed by \cite{Pac,Goo}, and has observational confirmations \cite{Gui}. 
 
 Assuming the thermalization does occur, we can estimate the fireball temperature from $E_c\sim l_\gamma ^3T_f^4$, giving
 \beq \label{Tf}
T_f \sim 1{\rm GeV}~ B_{-9}^{-1}\eta_{14}^{-3/2}\left( \frac{\gamma}{\gamma _c} \right)^{3/2}.
\eeq
The cusp event really produces an inhomogeneous fireball which is hotter and moves faster (with larger Lorentz factors)  on the inside than on the outside. But for the sake of simplicity, we will just speak of separate fireballs with different characteristic Lorentz factors $\gamma$.

Obviously, a necessary condition of thermalization is that the fireball be optically thick. Similar to \cite{Pac,Goo}, this requires $T_f\gtrsim 0.1$MeV, giving
\beq
\frac{\gamma}{\gamma _c}\gtrsim  2\times 10^{-3}~ B_{-9}^{2/3}\eta_{14}.
\eeq

One can argue that $T_f\gtrsim 0.1$MeV is also a sufficient condition of thermalization. This is because in the near zone of the EM pulse (at distances $\lesssim l_\gamma$ in the fireball rest frame) the electric and magnetic fields generated by the string, $E\sim B\sim T_f^2$ are generally not perpendicular to each other, with both field invariants of the same order. Then, for super-Schwinger fields, that is for $T_f\gtrsim  e^{-1/2}m_e\sim 2$ MeV, copious pair production does make the fireball optically thick. 

At somewhat lower effective temperatures $T_f$, Schwinger effect does not operate, but the tree-level QED processes are still expected to produce an avalanche of pairs, similar to what happens in pulsars \cite{Rud}: an accelerated electron emits gamma rays which produce pairs on the background field, pairs are then accelerated, etc... Quasi-thermalization is already seen in Crab-like pulsars,  even though the equivalent temperature, as measured at the light cylinder, is only about 1keV.  

A pulsar-like avalanche requires that curvature radiation photons be energetic enough to pair produce on the large-scale EM field. Consider then, closely following \cite{Rud}, a seed electron in a generic EM field $E\sim B\sim T_f^2$ with characteristic length scale $r\sim l_\gamma$. The electron quickly synchrotron cools, but as there is generically an electric field component along the magnetic field, the electron will keep moving relativistically along a curved magnetic field line, with curvature radius $\sim r$.  The terminal Lorentz factor of the electron, $\gamma _e$,  is determined by the balance between the electric field acceleration and the curvature radiation damping: 
\beq
eE\sim  \frac{e^2\gamma_e ^4}{r^2}.
\eeq
The electron, as it moves with Lorentz factor $\gamma_e$, emits curvature photons of energy
\beq
\epsilon \sim \frac{\gamma_e^3}{r}.
\eeq
These photons pair produce on the background field $B$, provided
\beq
\epsilon B\gtrsim 0.1\frac{m_e^3}{e},
\eeq
giving the avalanche criterion
\beq
T_f\gtrsim 1{\rm keV}~B_{-9}^{-2/17}\eta_{14}^{-6/17},
\eeq
which is always satisfied by a large margin if we use the proposed thermalization requirement $T_f\gtrsim 0.1$MeV.  We note also that particle ejected from the string can provide an additional mechanism for thermalizing fireballs with sub-Schwinger fields.

For effective temperatures somewhat smaller than 0.1 MeV, thermalization does not occur, as the plasma, if  thermalized, would have become optically thin. Still a very rough, but physically well motivated, estimate of the  emission is possible in some cases. This happens when the string parameter $\eta$, the ISM field $B$, and the fireball parameter $\gamma$ are such that the resulting EM field amplitude and length scale emulate the corresponding pulsar parameters as measured at the light cylinder. Assuming, as appears likely, that pulsar emission is mostly produced at the light cylinder, one can then expect that string loop cusps will produce similar pulsar-like emission. We will not pursue this anymore in this paper, limiting ourselves only to the "cleaner" cases with $T_f\gtrsim 0.1$MeV.

For even smaller $\gamma$ parameters, that is for larger beaming angles, a particle avalanche is not expected to develop. A solid angle $\sim \gamma ^{-2}$ is then illuminated by a pure-EM pulse. The EM pulse has very high energy density; it  sweeps up the ISM plasma, and some conversion of the pulse energy into high-energy photons (synchrotron and inverse Compton) must occur. It has even been suggested that the lower-$\gamma$ EM pulses from strings are responsible for GRBs \cite{BHV}.  We will not consider the low-$\gamma$ cases here because we do not have a full picture of the EM pulse interaction with the ISM. (Exactly for this  reason,  \cite{BHV} do not predict the photon energy in their GRB-from-strings scenario.)

\section{The  Fireball Evolution}\label{FE}

In what follows we assume that fireballs with effective temperatures $T_f\gtrsim 0.1$MeV do thermalize at about the temperature $T_f$. A distant observer will then see photons  of characteristic energy $\gamma T_f$ (and neutrinos with nearly the same characteristic temperature, if $T_f$ is high enough to produce thermalized neutrinos). Here the $\gamma$ factor simply accounts for the fireball frame motion with respect to the observer. 

Why $T_f$ gives the photon energy in the fireball rest frame, and why only the massless particles are released has been explained in  \cite{Pac,Goo}. As the fireball expands, the plasma temperature, of course, drops. At the same time, the plasma is radially accelerated to high Lorentz factors, and in exactly the same proportion as the temperature drop, so that the characteristic energy of a massless particle remains unchanged. The fireball becomes a spherical shell with a growing radius, but with the shell thickness close to the initial fireball radius. The photon distribution function evolves as if photons were not constantly scattered, produced and annihilated, although the photon-electron interactions truly terminate only when the temperature drops to about $0.1$MeV. At this temperature, the pair density is negligibly small, explaining why the fireball energy goes almost solely into photons (and possibly neutrinos), with only a negligible fraction given to leftover electrons/positrons and protons/antiprotons.

We have described fireball expansion into vacuum, while it actually expands into an ISM wind, blowing with Lorentz factor $\gamma$ in the fireball frame. One can check, however, that the swept-up protons are strongly energetically subdominant at the photon release temperature (negligible baryon loading, in the GRB terminology). This is intuitively clear, because the fireball proper size is initially $\lesssim 1$cm.

\section{Observables} \label{Obs}

 Three observationally promising scenarios will be considered in turn:

 (i) The entire energy is released into the ISM as photons of energy $\sim \gamma T_f$.  Some of these photons reach Earth.

(ii) An order unity  fraction of the fireball energy is converted into neutrinos of energy $\sim \gamma T_f$. These neutrinos freely propagate to Earth.

(iii) Almost all of the initial $\sim \gamma T_f$ photons are converted into cascade photons, producing an observable diffuse background.

\subsection{TeV Bursts}

By Eq.(\ref{Tf}), the fireball thermalizes (meaning, by our assumption, that the fireball has effective temperatures $T_f\gtrsim 0.1$MeV) provided
\beq
\gamma \gtrsim \gamma _B\equiv 4\times 10^7~B_{-9}^{-1/3}.
\eeq
This will inject photons of energy 
\beq\label{ep}
\epsilon \sim 4{\rm TeV}~B_{-9}^{-1/3}\left( \frac{\gamma}{\gamma _B} \right)^{5/2}.
\eeq
For photon energy $\epsilon\sim $few TeV, the photon mean free path (in the infrared background) can be crudely approximated \cite{Dom} by 
\beq
\lambda \sim 1{\rm Gpc}~\left( \frac{\epsilon}{1 {\rm TeV}} \right)^{-1}.
\eeq
The characteristic wait time for a burst of such photons is 
\beq
t\sim \frac{\gamma ^2t_c}{N}\left( \frac{3{\rm Gpc}}{\lambda}\right)^3,
\eeq
or
\beq\label{tp}
t\sim 3\times 10^{10}{\rm y}~B_{-9}^{-5/3}\eta_{14}^{5}\left( \frac{\gamma}{\gamma _B} \right)^{19/2}.
\eeq
The fluence (here defined as the number of photons per unit area) is 
\beq 
f\sim \frac{\gamma ^3E_c}{\epsilon \lambda ^2},
\eeq
or
\beq\label{fp}
f\sim 5\times 10^{9}\frac{1}{{\rm m}^2}~B_{-9}^{2/3}\eta_{14}^{6}\left( \frac{\gamma}{\gamma _B} \right)^{11/2}.
\eeq

To illustrate, take $B\sim 1$nG and $\eta \sim 10^{12}$GeV. Consider only the barely thermalized fireballs, with $\gamma \sim \gamma _B$. Then the photon energy, Eq.(\ref{ep}), the wait time, Eq.(\ref{tp}), and the fluence, Eq.(\ref{fp}), are
\beq
\epsilon \sim 4{\rm TeV},~~~~t\sim 3{\rm y},~~~~f\sim 0.005\frac{1}{{\rm m}^2}.
\eeq
Effective area of an all-sky TeV detector HAWC reaches $10^5{\rm m}^2$ at these energies. One can then get a (never seen before) many-photon event every few years.

A slight modification of this scenario is also worth mentioning. Individual sources, like the giant lobes of the nearby  radio galaxy Cen A, can become visible. A sufficiently high event rate is achievable for small enough symmetry breaking scale $\eta$. To illustrate, consider a constantly monitored object of size $r$ ($\sim 1$ Mpc) at a distance $d$ ($\sim 3$ Mpc) with magnetic field $B$ ($\sim 1\mu$G). If $\eta _{14}<<1$, there are many loops within a sphere of radius $r$, and an estimate along the same lines as above gives the event wait time, photon energy, and fluence
\beq
\epsilon \sim 400{\rm GeV},~~~~t\sim 5{\rm y}~\eta_{-11}^5,~~~~f\sim 0.3\frac{1}{{\rm m}^2}~\eta_{-11}^6.
\eeq
Effective area of HAWC at about 300 GeV is about  $10^3{\rm m}^2$, but Cen A is too far south for HAWC. Effective area of HESS at about 300 GeV is about  $10^5{\rm m}^2$, but HESS has a small field of view and observed Cen A for only about 100 hours. 

\subsection{Ultra-High-Energy Neutrino Bursts}

Neutrinos are produced and thermalized if the fireball is optically thick to electron-positron annihilation into neutrinos,
\beq
\sigma nr\sim (G_F^2T_f^2)T_f^3l_\gamma\gtrsim 1,
\eeq
or
\beq
\gamma \gtrsim \gamma _\nu\equiv 5\times 10^{9}~B_{-9}^{-5/11}\eta_{14}^{-4/11}, ~~~B_{-9}\lesssim 10~\eta_{14}^{-7/6},
\eeq
where the last inequality ensures that $\gamma _\nu$ does not exceed the maximal possible $\gamma =\gamma _c$.
Bursts of neutrinos of energy
\beq
\epsilon_\nu \sim  6\times 10^{17}{\rm eV}~B_{-9}^{-7/11}\eta_{14}^{-10/11}\left( \frac{\gamma}{\gamma _\nu} \right)^{5/2},
\eeq
are observed with the wait time
\beq
t_\nu \sim 3\times 10^{11}{\rm y}~B_{-9}^{-10/11}\eta_{14}^{47/11}\left( \frac{\gamma}{\gamma _\nu} \right)^{2},
\eeq
and fluences
\beq
f_\nu\sim 3\times 10^{14}\frac{1}{{\rm km}^2}~B_{-9}^{14/11}\eta_{14}^{64/11}\left( \frac{\gamma}{\gamma _\nu} \right)^{1/2}.
\eeq

To illustrate,  take $B\sim 1$nG and $\eta \sim 4\times 10^{11}$GeV. Consider only the barely $\nu$-thermalized fireballs, with $\gamma \sim \gamma _\nu$. Then the photon energy, the wait time, and the fluence are
\beq
\epsilon \sim 10^{20}{\rm eV},~~~~t\sim 15{\rm y},~~~~f\sim 3\frac{1}{{\rm km}^2},
\eeq
giving a multi-neutrino event in a detector with $\sim {\rm km}^2$ effective area.

\subsection{Diffuse gamma-ray background} 

Most energy (for a CMB-frame observer) is coming from the most compact part of the fireball, with $\gamma \sim \gamma _c$. These fireballs produce very narrowly beamed bursts of photons with characteristic energy
\beq\label{ep0}
\epsilon_0 \sim  2\times 10^{19}{\rm eV}~B_{-9}^{-2}\eta_{14}^{-5/2}.
\eeq
In about 10 Mpc, these ultra-high-energy photons produce an electron-positron pair on the CMB (or other radio background) \cite{Set}.

The electron and the positron have the same characteristic energy $\sim \epsilon_0$. These particles inverse-Compton scatter, pair produce, and emit synchrotron radiation. The synchrotron dominates roughly for
\beq
B_{-9}\gtrsim \left( \frac{\epsilon_0}{10^{19} {\rm eV}} \right)^{-1/2},
\eeq
which gives, using Eq.(\ref{ep0}), a $B$-independent criterion for the synchrotron domination
\beq
\eta _{14}\lesssim 1.
\eeq

If the synchrotron losses are the dominant ones, most initial energy gets converted into second  generation photons, with characteristic energy
\beq
\epsilon \sim  4{\rm GeV}~B_{-9}^{-3}\eta_{14}^{-5}.
\eeq
For $\epsilon \lesssim 100$GeV, the second generation photons freely propagate to Earth. What we see at Earth is not a burst, but a diffuse background. This is because synchrotron-emitting electrons and positrons were also deflected by the magnetic field. This spreads the cascade in angle and leads to large time delays. The diffuse gamma-ray flux is 
\beq 
F \sim  10^{-8}\frac{\rm erg}{\rm cm ^2\cdot s}~ B_{-9}.
\eeq

To illustrate,  take $B\sim 1$nG and $\eta \sim 10^{14}$GeV. Then the photon energy and flux are
\beq
\epsilon \sim 4{\rm GeV},~~~F \sim  10^{-8}\frac{\rm erg}{\rm cm ^2\cdot s}.
\eeq
This is in very strong conflict with observations. Not only is the actual background at 4GeV smaller ($\sim 3\times 10^{-10}\frac{\rm erg}{\rm cm ^2\cdot s\cdot sr}$, \cite{Fer}), almost all of the observed background seems to have a mundane (astrophysical) explanation \cite{Hoo}.

\section{Conclusions}

We have argued that the energy discharge from cusps of superconducting strings occurs in the form of thermalized relativistic fireballs.  The large Lorentz factor $\gamma$ of the cusp boosts the energies of thermal particles and collimates them in a narrow beam with opening angle $\gamma^{-1}$.  As the fireball expands and cools, practically all massive particle-antiparticle pairs annihilate, so only photons and neutrinos can reach the observer.  

The predicted neutrino and photon fluxes are not sensitive to string microphysics and depend only on the string energy scale $\eta$ and on the magnitude $B$ and the volume filling factor $f_B$ of the cosmic magnetic fields.  Since the magnitude and the distribution of intergalactic magnetic fields are highly uncertain, we could not derive any firm bounds on the string parameter $\eta$.  However, with plausible assumptions about $B$ and $f_B$, we showed that strings with $\eta \sim 10^{10}-10^{15} ~GeV$ can produce observable effects with unique signatures.  These include TeV photon bursts (coincident multiple hits of the detector area by TeV photons),  UHE neutrino bursts, and TeV photon bursts from individual astrophysical sources.\footnote{The possibility of such effects was first indicated in Ref.\cite{BOSV}, but their predicted rate and fluxes are different in the fireball scenario.}  Both neutrino and photon bursts from cusps can be accompanied by potentially detectable bursts of gravitational radiation \cite{BHV}. The diffuse gamma-ray background from strings does not have unique signatures but provides stringent combined constraints on the symmetry breaking scale and the ISM magnetic field.

\section*{Acknowledgements}

A.V. is grateful to Venya Berezinsky and to Ken Olum for useful discussions and to Gregory Gabadadze for his hospitality at New York University, where this work was initiated.  This work was supported in part by the National Science Foundation under grant PHY-1518742 (A.V.).


\begin{thebibliography}{99}

\bibitem{Book}
A. Vilenkin and E. P. S. Shellard, {\it Cosmic Strings and Other Topological Defects}
(Cambridge University Press, Cambridge, 2000).

\bibitem{Witten}
E. Witten, "Superconducting cosmic strings", 
Nucl. Phys. B249, 557 (1985).

\bibitem{VV87} 
  A.~Vilenkin and T.~Vachaspati,
  ``Electromagnetic Radiation from Superconducting Cosmic Strings,''
  Phys.\ Rev.\ Lett.\  {\bf 58}, 1041 (1987).

\bibitem{Spergel87} 
  D.~N.~Spergel, T.~Piran and J.~Goodman,
  ``Dynamics of Superconducting Cosmic Strings,''
  Nucl.\ Phys.\ B {\bf 291}, 847 (1987).

\bibitem{Babul87} 
  A.~Babul, B.~Paczynski and D.~Spergel,
  ``Gamma-ray bursts from superconducting cosmic strings at large redshifts,''
  Astrophys.\ J.\  {\bf 316}, L49 (1987).

\bibitem{Barr87} 
  S.~M.~Barr and A.~M.~Matheson,
  ``Limiting Currents in Fermionic Superconducting Strings,''
  Phys.\ Lett.\ B {\bf 198}, 146 (1987).

\bibitem{BOSV} 
  V.~Berezinsky, K.~D.~Olum, E.~Sabancilar and A.~Vilenkin,
  ``UHE neutrinos from superconducting cosmic strings,''
  Phys.\ Rev.\ D {\bf 80}, 023014 (2009)
 
\bibitem{Tanmay} 
  T.~Vachaspati,
  ``Cosmic Rays from Cosmic Strings with Condensates,''
  Phys.\ Rev.\ D {\bf 81}, 043531 (2010)
    [arXiv:0911.2655 [astro-ph.CO]]. 
 
 \bibitem{BPOS} 
  J.~J.~Blanco-Pillado, K.~D.~Olum and B.~Shlaer,
  ``The number of cosmic string loops,''
  Phys.\ Rev.\ D {\bf 89}, no. 2, 023512 (2014)  

\bibitem{BPO16} 
  J.~J.~Blanco-Pillado, K.~D.~Olum and B.~Shlaer,
  ``Cosmic string loop shapes,''
  Phys.\ Rev.\ D {\bf 92}, no. 6, 063528 (2015)
    [arXiv:1508.02693 [astro-ph.CO]]. 

\bibitem{Rubinstein} 
  V.~Berezinsky and H.~R.~Rubinstein,
  ``Evolution and Radiation From Superconducting Cosmic Strings,''
  Nucl.\ Phys.\ B {\bf 323}, 95 (1989).
   
\bibitem{BPOV} 
  J.~J.~Blanco-Pillado, K.~D.~Olum and A.~Vilenkin,
  ``Dynamics of superconducting strings with chiral currents,''
  Phys.\ Rev.\ D {\bf 63}, 103513 (2001)  

\bibitem{BPO}
  J.~J.~Blanco-Pillado and K.~D.~Olum,
  ``Electromagnetic radiation from superconducting string cusps,''
  Nucl.\ Phys.\ B {\bf 599} (2001) 435
    [astro-ph/0008297].
  
 \bibitem{BHV} 
  V.~Berezinsky, B.~Hnatyk and A.~Vilenkin,
  ``Gamma-ray bursts from superconducting cosmic strings,''
  Phys.\ Rev.\ D {\bf 64}, 043004 (2001)
    [astro-ph/0102366].

\bibitem{Goo} 
J. Goodman, 
"Are gamma-ray bursts optically thick?",
ApJ 308, L47 (1986)

\bibitem{Pac} 
B. Paczynski,
"Gamma-ray bursters at cosmological distances",
ApJ 308, L43 (1986)

\bibitem{Gui}
S. Guiriec et al.,
"GRB 131014A: A Laboratory for Studying the Thermal-like and Non-thermal Emissions in Gamma-Ray Bursts...",
ApJ 814, 10 (2015)

\bibitem{Rud}
M. A. Ruderman, P.G. Sutherland, 
"Theory of pulsars - Polar caps, sparks, and coherent microwave radiation",
ApJ  196, 51 (1975)

\bibitem{Dom}
A. Dominguez et. al.,
"Extragalactic background light inferred from AEGIS galaxy-SED-type fractions",
Mon. Not. R. Astron. Soc. 410, 2556 (2011)

\bibitem{Set}
M Settimo, M. De Domenico,
"Propagation of extragalactic photons at ultra-high energy with the EleCa code",
 Astroparticle Physics 62, 92 (2014)
 
 \bibitem{Fer}
 M. Ackermann et al.,
 "The Spectrum of Isotropic Diffuse Gamma-Ray Emission between 100 MeV and 820 GeV",
 ApJ 799, 86 (2015)
 
 \bibitem{Hoo}
 
 D. Hooper, T. Linden, A. Lopez,
 "Radio Galaxies Dominate the High-Energy Diffuse Gamma-Ray Background",
 arXiv 1604.08585 (2016)
 
 
\end{thebibliography}
\end{document}